\documentclass[manuscript]{elsarticle}
\usepackage[gen]{eurosym}
\usepackage{graphicx}
\usepackage{amsmath}
\usepackage[margin=1in,marginpar=1in]{geometry}
\usepackage{tabularx}
	\newcolumntype{Z}{>{\centering\arraybackslash}p}
\usepackage{setspace}
\usepackage[usenames, dvipsnames]{color}
\usepackage[normalem]{ulem}

\begin{document}
	
\begin{frontmatter}
		
\title{Improving optimal control of grid-connected lithium-ion batteries through more accurate battery and degradation modelling}
\author[OU,EV,VIT]{Jorn M. Reniers} 
	\ead{jorn.reniers@eng.ox.ac.uk}
\author[EV,VIT]{Grietus Mulder} 
	\ead{grietus.mulder@energyville.be}
\author[OU]{Sina Ober-Bl$ \mathrm{\ddot o}$baum} 
	\ead{sina.ober-blobaum@eng.ox.ac.uk}
\author[OU]{David A. Howey \corref{cor1}} 
	\ead{david.howey@eng.ox.ac.uk}
	\ead[url]{http://epg.eng.ox.ac.uk/}

\address[OU] {Department of Engineering Science, University of Oxford, OX1 3PJ, Oxford, UK}
\address[EV] {EnergyVille, Thor Park 8310, 3600, Genk, Belgium}
\address[VIT] {VITO, Boeretang 200, 2400 Mol, Belgium}

\cortext[cor1]{Corresponding author}

\begin{abstract}
	The increased deployment of intermittent renewable energy generators opens up opportunities for grid-connected energy storage. Batteries offer significant flexibility but are relatively expensive at present. Battery lifetime is a key factor in the business case, and it depends on usage, but most techno-economic analyses do not account for this. For the first time, this paper quantifies the annual benefits of grid-connected batteries including realistic physical dynamics and nonlinear electrochemical degradation. Three lithium-ion battery models of increasing realism are formulated, and the predicted degradation of each is compared with a large-scale experimental degradation data set (Mat4Bat). A respective improvement in RMS capacity prediction error from 11\% to 5\% is found by increasing the model accuracy. The three models are then used within an optimal control algorithm to perform price arbitrage over one year, including degradation. Results show that the revenue can be increased substantially while degradation can be reduced by using more realistic models. The estimated best case profit using a sophisticated model is a 175\% improvement compared with the simplest model. This illustrates that using a simplistic battery model in a techno-economic assessment of grid-connected batteries might substantially underestimate the business case and lead to erroneous conclusions.
	
	\textit{keywords}: Lithium Ion, Degradation, Grid-connected Battery, Optimal control, Battery ageing, Battery modelling
\end{abstract}

\end{frontmatter}

\newpage 

\section{Introduction}


Challenges for the electricity system arise due to increasing deployment of intermittent renewable energy sources \cite{Doetsch2015}. For example, balancing production and demand becomes more difficult, grid inertia decreases, and distribution grids become more congested. As part of a broad portfolio of possible solutions, battery energy storage provides a flexible option to address many of these problems. However, the lifetime of a battery in terms of capacity and power capability strongly impacts the profitability of battery storage \cite{Battke2013}. The lifetime of a lithium-ion (li-ion) battery depends on how it is used because there are multiple degradation mechanisms, each influenced by different usage patterns \cite{Vetter2005}. 

Many previous economic assessments of storage have included a battery degradation model, usually an empirical correlation based on fitting of measured degradation tests e.g.\ \cite{Battke2013,Swierczynski2014,Perez2016a,Uddin2017}. Although empirical models can provide valuable insight, they should be used with caution \cite{Jokar2016,Bizeray2015}. They are based on a limited number of test conditions and do not necessarily apply to other load profiles, risking extrapolation without theoretical basis. Furthermore, battery characteristics change as batteries age, which is often not taken into account. Finally, empirical models only apply to the exact type of cell for which they have been developed. 

A few researchers have used electrochemical models to address these issues, and initial results are promising: a more intelligent battery utilisation informed by a physical model could decrease battery degradation. Lawder et al.\ \cite{Lawder2015} compared battery models for a simple micro-grid application (ignoring degradation) and noted how accumulated errors in equivalent circuit battery models led to substantial discrepancies between the simulated and real state of charge (SoC). Multiple researchers, e.g.\ \cite{Pathak2017,Perez2016}, used electrochemical battery models to optimise charging profiles, increasing battery life. Others have used electrochemical battery degradation models without optimisation to analyse specific case studies\ \cite{Lee2017, Patsios2016,Weißhar2016}.

These cases studies showed the potential for using electrochemical battery models to improve the lifetime and therefore the economic impact of grid-connected batteries. However, to our best knowledge, due to the complexity of these nonlinear battery models, they have not been used for optimisation over a long time horizon (e.g. a year), yet this is required to truly quantify their performance. Therefore, this paper aims to identify the economic performance gains achievable by using a nonlinear, electrochemical battery model, including realistic dynamics and degradation, in an economic optimisation for a realistic grid application over a full year of data.

\section{Nomenclature}
\begin{center}
	\begin{tabular}{p{1.5cm} p{12.5cm} p{2cm}}
			$\alpha$ 	& Degradation parameter in equivalent circuit model 	& 			\\
			$\beta$ 	& Degradation parameter in equivalent circuit model 	& 			\\
			$\lambda(t)$& Wholesale electricity price at time \textit{t}			& $\mathrm{\euro{}\ (Wh)^{-1}}$ \\ 
			$\lambda_{ \text{degr,Wh}}$& Cost of battery energy degradation						& $\mathrm{\euro{}\ (Wh)^{-1}}$ \\
			$\lambda_{ \text{degr,Ah}}$& Cost of battery charge degradation						& $\mathrm{\euro{}\ (Ah)^{-1}}$ \\
			$\lambda_{ \text{degr,LLI}}$& Cost of lost cyclable lithium							& $\mathrm{\euro{}\ (Ah)^{-1}}$ \\
			$ C $  		& Battery degradation cost 									& $\mathrm{\euro{}}$ \\
			$ C_p$ 	& Parallel capacitor in equivalent circuit model 		& $\mathrm{F}$	\\
			$ c_i(r,t)$ & Lithium concentration in electrode \textit{i} at radius \textit{r} and time \textit{t} in single particle model									& $\mathrm{mol\ m^{-3}}$ \\
			$ c_{i}^{ \text{max}}$ & Maximum lithium concentration in electrode \textit{i} 	& $\mathrm{mol\ m^{-3}}$ \\
			$ E_{ \text{Wh}}$ 	& Battery energy capacity 									& $\mathrm{Wh}$ 		\\
			$ E_{ \text{Ah}}$ 	& Battery charge capacity									& $\mathrm{Ah}$ 		\\
			$ E_{ \text{lost,Wh}}$ & Lost battery energy capacity							& $\mathrm{Wh}$ 		\\
			$ E_{ \text{lost,Wh}}$ & Lost battery charge capacity 							& $\mathrm{Ah}$ 		\\
			$ f $ 		& Battery state space model	 								& 	 		\\
			$ g $ 		& Constraint function 										&  			\\
			$I(t)$ 	& Battery current at time \textit{t} 						& $\mathrm{A}$ 		\\
			$I_r(t)$ 	& Current through the parallel resistor in the equivalent circuit model at time \textit{t}																& $\mathrm{A}$ 		\\
			$L(T_{ \text{end}})$& Lost cyclable lithium at the end of the simulation time	& $\mathrm{Ah}$ 		\\
			$N$ 		& Number of cells in the battery							& $\mathrm{-}$ 		\\
			$OCV$ 		& Open circuit voltage in the equivalent circuit model		& $\mathrm{V}$ 		\\
			$P(t)$ 		& Power to/from the battery at time \textit{t}				& $\mathrm{W}$ 		\\
			$R$ 		& Revenue per unit of time									& $\mathrm{\euro{}\ h^{-1}}$ \\
			$R_p$ 		& Parallel resistor in the equivalent circuit model			& $\mathrm{\Omega}$ 	\\
			$R_s$ 		& Series resistor in the equivalent circuit model			& $\mathrm{\Omega}$ 	\\
			$T_{ \text{end}}$ 	& Total simulation time										& $\mathrm{h}$ 		\\
			$u(t)$ 		& Control variables at time \textit{t}						&  			\\
			$u_n(t)$	& Control variables in optimisation \textit{n} at time \textit{t} & 	\\
			$U^{ \text{opt}}(t)$& Optimal control variables at time \textit{t} 				& \\
			$U_n^{ \text{opt}}(t)$& Optimal control variables in optimisation \textit{n} at time \textit{t} & \\
			$V(t)$ 		& Battery voltage at time \textit{t}						& $\mathrm{V}$ 		\\
			$V_{ \text{mean}}$	& Mean battery voltage										& $\mathrm{V}$ 		\\
			$x(t)$ 		& State variables at time \textit{t}						& 			\\
			$x_n(t)$ 	& State variables at time \textit{t} in optimisation \textit{n}&        \\
			$z(t)$ 	    & State of charge at time \textit{t}						& $\mathrm{-}$ 	
	\end{tabular}
\end{center}

\section{Methods}
	\subsection{Problem setup}
\begin{figure}
			\centering
			\includegraphics[width=9cm]{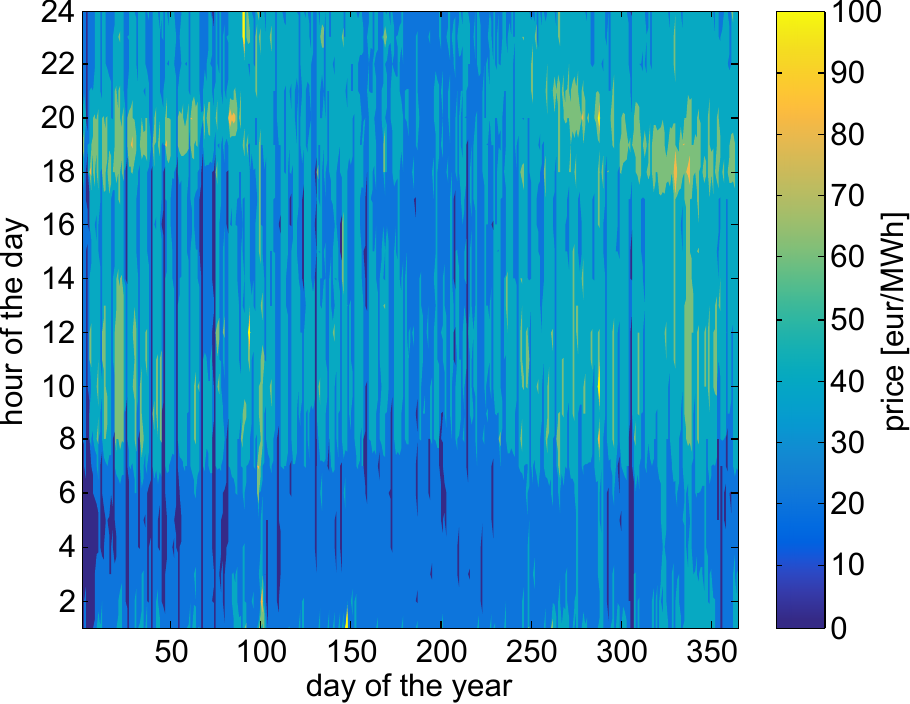}
			\caption{Wholesale price on the day-ahead market in Belgium in 2014 \cite{Belpex2016} }
			\label{Pwholesale}
		\end{figure}

		In this simulation study, a lithium-ion battery was used for price arbitrage. In other words, revenue was made by buying energy on the wholesale market when prices were low, charging the battery, and then selling energy on the market at higher prices at a later point, discharging the battery. However, usage of the battery also resulted in capacity fade. The task of a battery operator who wishes to exploit this market is to identify a load profile which maximises revenue and minimises lost capacity. For price data we used the wholesale price of the Belgian day-ahead electricity market in 2014, shown in Figure \ref{Pwholesale}, where the colour indicates the price at each hour (y-axis) of each day (x-axis). The price was assumed to be known perfectly, leading to a deterministic optimisation problem, described below. 
		
		A generic optimal control formulation was used to describe this problem mathematically (\ref{objectiveFunc}-\ref{constraint}). The revenue per unit of time \textit{R} and the degradation cost \textit{C} are both a function of the control variable \textit{u}(\textit{t}) and the state variable  \textit{x}(\textit{t}). A state-space model \textit{f} for the battery relates the state variables to the control variables and initial states. Depending on the battery state space model, a different physical meaning is given to the control and state variables. Other constraints such as the voltage limits of the battery, were incorporated into the constraint function \textit{g}.
		
		\begin{equation}
			\text{argmax}_u \left[ - C \left(u \left(t \right),x \left(t \right) \right) + \int_{0}^{T_{ \text{end}}}R \left(u \left(t \right),x \left(t \right) \right)dt \right]
			\label{objectiveFunc}
		\end{equation}
		
		subject  to
		\begin{equation}
			\frac{dx}{dt}=f(u(t),x(t))
			\label{stateSpaceModel}
		\end{equation}
		\begin{equation}
			g(u(t),x(t)) \geq 0
			\label{constraint}
		\end{equation}
		
	\subsection{Battery models} \label{batterymodels}
\begin{figure}
			\centering
			\includegraphics[width=14cm]{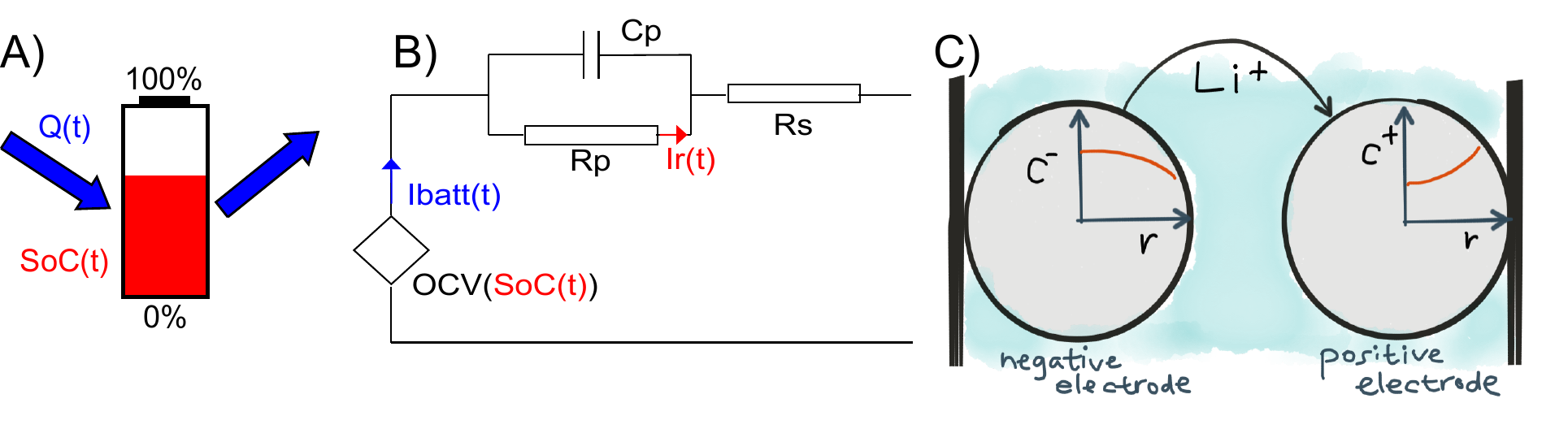}
			\caption{Battery models. A) bucket model; B) equivalent circuit model; C) single particle model}
			\label{BatteryModels}
		\end{figure}

	One cell with a capacity of 2.7 Ah was simulated with three commonly used battery models shown in Figure \ref{BatteryModels}, each employing a different degradation model. A detailed presentation and discussion of the assumptions for each battery model can be found in literature \cite{Jokar2016, Seaman2014a, Ahmadian2018}. A battery pack with 750 cells was modelled by multiplying the revenue and degradation cost for this cell by the number of cells. This assumes that all cells behave equally, which is only true in a high quality, well balanced and well managed pack or if every cell is controlled individually \cite{Frost2017}.
		
		\paragraph{Bucket model}The simplest model considers a battery as a repository for energy, as shown in Figure \ref{BatteryModels}a, much like a fuel tank. The control variable \textit{u} is the power flow to/from the battery \textit{P}. The only state variable \textit{x} is the state of charge \textit{z} and hence the battery state space model \textit{f} consists of one equation (\ref{BM_f}), with two constraints \textit{g} (\ref{BM_g}). The revenue per unit of time \textit{R} is the product of the power and the price (\ref{BM_R}). The battery was assumed to reach its end of life after 8000 full cycles when 20\% of the capacity has been lost, leading to a degradation factor of $1.25 E^{-5}$ per energy throughput (\ref{BM_degrad}). To ensure a constant power level as long as the price is constant, a small penalty for the maximum power level was introduced. The degradation cost \textit{C} was the product of the lost capacity and the cost of degradation (\ref{BM_C}). 
		
		\begin{equation}
			\frac{dz(t)}{dt}=\frac{P(t)}{E_{ \text{Wh}}}
			\label{BM_f}
		\end{equation}
		\begin{equation}
			0 \leq z(t) \leq 1
			\label{BM_g}
		\end{equation}
		\begin{equation}
			R \left(u \left(t \right),x \left(t \right) \right)=P(t)\lambda(t)N
			\label{BM_R}
		\end{equation}
		\begin{equation}
			E_{ \text{lost,Wh}}=2.15 \cdot 10^{-4} \cdot max|P(t)| + 1.25 \cdot 10^{-5} \cdot\int_{0}^{T_{ \text{end}}}|P(t)|dt
			\label{BM_degrad}
		\end{equation}
		\begin{equation}
			C \left(u \left(t \right),x \left(t \right) \right) =E_{ \text{lost,Wh}}\lambda_{degrad,Wh}N
			\label{BM_C}
		\end{equation}
		
		\paragraph{Equivalent circuit model} In a more complex model, a battery was represented by the electrical circuit shown in Figure \ref{BatteryModels}b. The control variable \textit{u} is the current to/from the battery \textit{I}. There are two state variables \textit{x}: the state-of-charge \textit{z} and the current through the parallel resistor $I_r$. The state-space model \textit{f} consists of two equations (\ref{ECM_fsoc}-\ref{ECM_fir}) \cite{Plett2015}. The voltage may be calculated using Ohm’s law (\ref{ECM_V}). There are four constraints \textit{g} (\ref{ECM_gv}-\ref{ECM_gsoc}). The revenue per unit of time \textit{R}  is the product of the power and the price (\ref{ECM_R}). The estimated degradation is based on the empirical degradation formula by Schmalstieg et al.\ \cite{{Schmalstieg2014}} (\ref{ECM_degrad}), which is a function of the simulation time, the charge throughput and the charge capacity of the battery. Within this model, the variable $\alpha$ is a function of the mean voltage and the temperature, while $\beta$ is a function of the root mean square voltage and the average deviation from the mean SoC (given by $2\frac{\int |z_{ \text{mean}}-z(t)|dt}{T_{ \text{end}}}$). This SoC-deviation was used to approximate the ‘depth of discharge’ in the original equation by Schmalstieg et al.\ The degradation cost \textit{C} was calculated according to equation (\ref{ECM_C}).
		
		\begin{equation}
			\frac{dz(t)}{dt}=\frac{I(t)}{E_{ \text{Ah}}}
			\label{ECM_fsoc}
		\end{equation}
		\begin{equation}
			\frac{dI_r(t)}{dt}=\frac{1}{R_pC_p}I(t) - \frac{1}{R_pC_p}I_r(t)
			\label{ECM_fir}
		\end{equation}
		\begin{equation}
			V(t) = OCV \left(z \left(t \right) \right)-R_pI_r(t)-R_sI(t)
			\label{ECM_V}
		\end{equation}
		\begin{equation}
			2.7 \leq V(t) \leq 4.2
			\label{ECM_gv}
		\end{equation}
		\begin{equation}
			0 \leq z(t) \leq 1
			\label{ECM_gsoc}
		\end{equation}
		\begin{equation}
			R \left(u \left(t \right),x \left(t \right) \right)=I(t)V(t)\lambda(t)N
			\label{ECM_R}
		\end{equation}
		\begin{equation}
			E_{ \text{lost,Wh}}= \left ( \alpha \left( V_{ \text{mean}},T(t) \right) T_{ \text{end}}^{0.75} + \beta \left( V(t),z(t) \right) \sqrt{\int ( |I(t)|dt )} \right)E_{ \text{Ah}}
			\label{ECM_degrad}
		\end{equation}
		\begin{equation}
			C \left(u \left(t \right),x \left(t \right) \right)= E_{ \text{lost,Wh}}\lambda_{degrad,Ah}N
			\label{ECM_C}
		\end{equation}

		\paragraph{Single particle model (SPM)} The SPM \cite{Ning2004} is one of the simplest electrochemical battery models; electrolyte transport is ignored, and only electrode transport is modelled. Each electrode is represented by one sphere in which lithium ions diffuse according to Fick’s law of diffusion, as shown in Figure \ref{BatteryModels}c. Chebyshev collocation was used to discretise the spatial derivative of the diffusion equation \cite{Bizeray2015} in our implementation of the model. At the surface, li-ions react according to Butler-Volmer reaction kinetics. The SPM was extended with a lumped thermal model \cite{Guo2011} and with a parasitic side reaction consuming cyclable li-ions in order to grow a passivating layer (the solid electrolyte interphase, or SEI) on the graphite electrode, thereby decreasing the cell capacity \cite{Christensen2005}. The SEI layer grows both during rest and on cycling. Its growth is enhanced by high temperature, high state of charge and high power charge. The resulting model was similar to other single particle models used often by battery researchers to predict battery degradation, e.g.\ \cite{Patsios2016,Pinson2013,Ramadass2004}. The control variable \textit{u} is the current to/from the battery \textit{I}, and there are 13 state parameters \textit{x}: the lithium concentration at the 5 Chebyshev nodes in each electrode, the battery temperature, the thickness of the SEI layer and the amount of li-ions consumed by the growing SEI layer. The state space model equations \textit{f} are given in \ref{appendix_SPM}. There were four constraints \textit{g} (\ref{SPM_gc}-\ref{SPM_gv}). The revenue per unit of time \textit{R} was calculated similarly as for the other models (\ref{SPM_R}). The degradation cost \textit{C} was the product of the loss of li-ions at the end of the simulation time, the cost of lost lithium inventory and the number of cells (\ref{SPM_C}).
		
		\begin{equation}
			0 \leq c_i(r,t) \leq c_i^{ \text{max}}
			\label{SPM_gc}
		\end{equation}
		\begin{equation}
			2.7 \leq V(t) \leq 4.2
			\label{SPM_gv}
		\end{equation}
		\begin{equation}
			R \left(u \left(t \right),x \left(t \right) \right) =I(t)V(t)\lambda(t)N
			\label{SPM_R}
		\end{equation}
		\begin{equation}
			C \left(u \left(t \right),x \left(t \right) \right)=L(T_{ \text{end}})\lambda_{ \text{degr,LLI}}N
			\label{SPM_C}
		\end{equation}
		
		All three of the aforementioned battery models used a slightly different cost of battery degradation. The bucket model used the cost of battery energy degradation $\lambda_{ \text{degr,Wh}}$, which was set at 0.33 $\mathrm{\euro{}\  (Wh)^{-1}}$, i.e.\ about 450 \$ $\mathrm{(kWh)^{-1}}$ in 2014 \cite{Nykvist2015}. The equivalent circuit model used the charge degradation cost $\lambda_{ \text{degr,Ah}}$, which was set to 1.2 $\mathrm{\euro{}\  (Ah)^{-1}}$ by assuming an average voltage of 3.64 V. The single particle model used the cost of lost lithium inventory. It has been shown that the lost charge capacity is not the same as the lost lithium inventory in a real battery \cite{Birkl2017}. In our simulations with the single particle model, the lost lithium was calculated explicitly. The capacity was measured by simulating a full charge and discharge, similar to the way it is measured in experiments. The lost capacity is the difference between the result of this and the initial battery capacity. Comparing the lost capacity and lost lithium showed that they are very similar, probably because no loss of active material is included in our model here. Therefore the price of lost lithium inventory $\lambda_{ \text{degr,LLI}}$ was also set at 1.2  $\mathrm{\euro{}\  (Ah)^{-1}}$, where the ‘Ah’ is the lost lithium.
	
	\subsection{Optimisation algorithm}
		The bucket battery model is input-output linear, and hence long time periods can be simulated easily, explaining its widespread use. Standard linear solvers were used. 
		
		The equivalent circuit battery model is linear apart from the open circuit voltage curve \textit{OCV(z(t))} and the degradation cost \textit{C}. The nonlinear optimisation techniques described below for the single particle model worked for this battery model too, and due to the nonlinearity only being in the output equation, and the low order of the model, the optimisation was about 10 times faster than the single particle model (but over 1000 times slower than the bucket model).
		
		The single particle model is a higher order model consisting of nonlinearly coupled nonlinear partial differential equations and optimisation is more challenging. As a prerequisite, the model has to be formulated with the optimisation in mind. The spatial Chebyshev discretisation used (see section \ref{batterymodels}) reduced the number of state variables compared to the conventional technique of finite differences. Simultaneous time discretisation was performed by applying a time integration scheme to the battery model. A forward Euler time integration scheme was chosen leading to relatively simple equations and a sparse Jacobian. The maximum time step size was 5 seconds to ensure numerical stability. Multiple shooting \cite{Stoer2002} was then used to reduce the number of optimisation variables. Instead of having one vector of state variables per 5 seconds, the model only had one per 15 minutes. Starting from one vector of state variables, the state space model was integrated over 15 minutes, in steps of 5 seconds, and the resulting vector at the end of this period had to be the same as the next vector of state variables.
		
		This approach resulted in a set of (nonlinear) algebraic constraints linking the optimisation variables at different points in time. Then IPOPT \cite{Biegler2009}, a gradient-based nonlinear optimisation package, was used to maximise the profit. The limited-memory quasi-Newton method was used to approximate the Hessian of the Lagrangian. MUMPS \cite{Amestoy2006} and HSL \cite{Science&TechnologyFacilityCouncil2013} were used as linear solvers. Automatic differentiation performed by ADOL-C \cite{Walther2012} and its sparse drivers ColPack \cite{Gebremedhin2013} was used to calculate the derivatives needed by IPOPT. 
		
	\subsection{Sliding window optimisation} \label{slidingWindowParagraph}
\begin{figure}
			\centering
			\includegraphics[width=9cm]{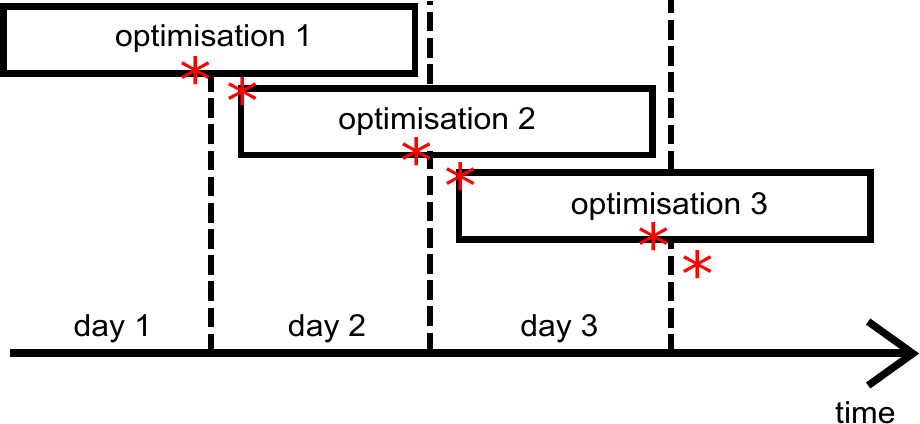}
			\caption{Sliding window approach to simulating long periods. Stars indicate the transition constraints}
			\label{slidingWindow}
		\end{figure}

		With the previous optimisation techniques, the single particle model could be optimised for periods of several weeks relatively fast: it took about 6 hours to optimise a period of one week at a time resolution of 5 seconds; involving 10752 optimisation variables. To simulate longer periods, a sliding window approach was adopted (\ref{slide_loop}-\ref{slide_transCon}). It was found that the size of the window affected the outcome only minimally beyond a size of one day. Therefore, a two day window was chosen as shown in Figure \ref{slidingWindow}. 
		
		First, the battery utilisation was optimised for days 1 and 2 (\ref{slide_f}-\ref{slide_constraint}). This yielded utilisation profiles for both days. The profile for day 1 was considered the ‘optimal’, while day 2 was included to account for end-effects (energy stored in the battery at the end of day 1 would have a value in the future) and hence the profile for day 2 was discarded (\ref{slide_Uopt}). Then, the battery state at the end of day 1 was taken as the start point for the second optimisation (\ref{slide_transCon}) as indicated by the stars on in Figure 3. The second optimisation optimised the battery utilisation of day 2 and 3, and so on.
		
		The approach is very similar to Model Predictive Control (MPC). The difference is that there are no measurements done to update the states after one optimisation window. Just like MPC, the sliding window approach will lead to suboptimal results because the entire time horizon is not considered.
		
		\begin{align}
			\label{slide_loop}
			\text{for }\ n&=0\ \text{to}\ 364  \\
				\label{slide_f}
				&U_n^{ \text{opt}}= \text{argmax}_{u_n} \left[ -C(u_n(t),x_n(t)) + \int_{24n}^{24(n+2)}R(u_n(t),x_n(t))dt \right] \\
				&\text{Subject\ to} \\
					\label{slide_stateSpaceModel}
					& \indent \indent \frac{dx_n}{dt}=f(u_n(t),x_n(t)) \\
					\label{slide_constraint}
					& \indent \indent g(u_n(t),x_n(t)) \geq 0 \\
					\label{slide_Uopt}
				&U^{ \text{opt}}(24n\ \text{to}\ 24(n+1)-1) = U_n^{ \text{opt}}(0\ to\ 23) \\
				\label{slide_transCon}
				&x_{n+1}(0)=x_n(24) 
		\end{align}
		
	\subsection{Post processing} \label{postProcess}
		The outcome of the optimisation is the optimal current at each point in time for each battery model. However, this current might be infeasible in a real battery of the same capacity because it would result in under- or over-charging. For example, the optimal profile for the bucket model typically includes a high power input current until the battery is fully charged. But in a real battery this would lead to overcharging due to diffusion delays, overpotentials, resistive voltage drops, etc. Therefore, the battery in this scenario had to be `oversized' to use the optimal profile, or alternatively the capacity of the battery in the model had to be decreased in order to introduce safety margins to guarantee a current that would be safe for a real battery. This second approach was followed. There is however an alternative control method that could be used, which is to hold the cell at a voltage limit if that limit is reached. As a comparison, \ref{appendix_postProcess} introduces this way to avoid the over-and under-voltages and discusses how this changes the quantitative results, but not our overall conclusions.
		
		Due to time constraints, in the absence of experimental tests, the single particle model, described previously, was used here as our `real battery' for validation purposes. First, the optimal current was given as the input to the SPM which calculated the battery voltage. If this voltage exceeded the safety limits, the optimal current was scaled-down. The battery started at 50\% SoC, so scaling down the current meant the battery stayed closer to 50\% SoC, avoiding under- and overcharging. The single particle model was then also used to estimate the `real' lost capacity corresponding to the optimal (down-scaled) currents. The capacity at a certain point in time was measured by simulating a full charge/discharge cycle at low current starting from the battery state at that point in time as explained in section \ref{batterymodels}. The decrease in capacity, multiplied by the cost of battery degradation $\lambda_{ \text{degr,Ah}}$, gave an estimate for the degradation cost.
		
\section{Results}
	\subsection{Accuracy of battery degradation models}
\begin{figure}
			\centering
			\includegraphics[width=17cm]{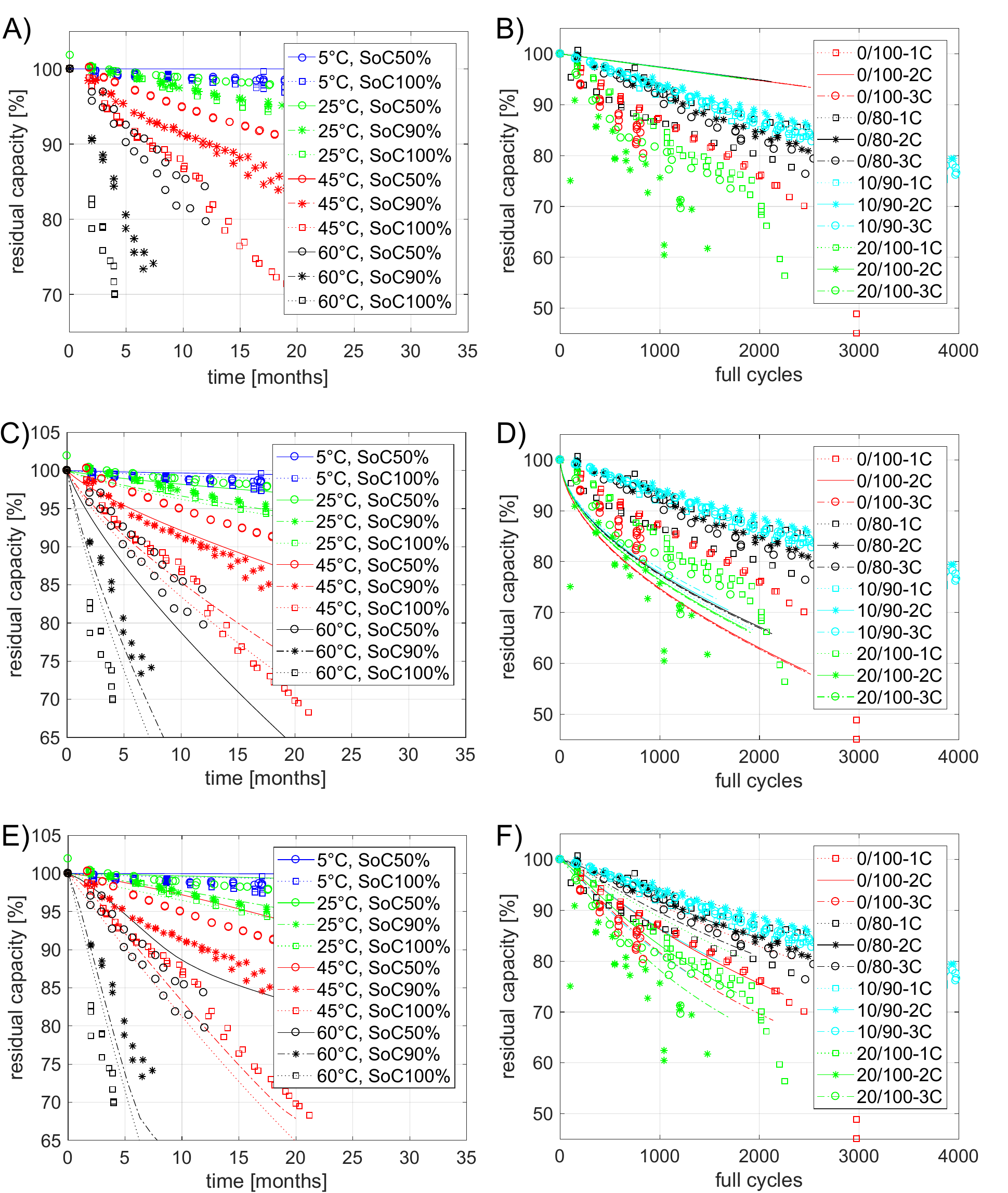}
			\caption{Measured (marker) and simulated (line) relative remaining capacity. Left: calendar ageing at various SoC and temperatures. Right: cycle ageing at various SoC windows and various charging currents, at a temperature of 45$^{\circ}$C. A) \& B) bucket model; C) \& D) equivalent circuit model; E) \& F) single particle model}
			\label{degradation}
		\end{figure}

		Within the Mat4Bat project \cite{EuropeanCommission2016}, multiple degradation experiments were performed on a Kokam 16 Ah NMC cell. Calendar ageing at various temperatures and SoCs was measured, as was cycle ageing in various SoC windows and at various charge currents (discharge was always at a 1C rate) at 45$^{\circ}$C. The three battery state space models were integrated over time with the current cycles from the Mat4Bat experiments to simulate equivalent degradation. This simulated capacity degradation (lines) is compared with the experimental data (markers) in Figure \ref{degradation}. The x-axis of the cycle ageing graphs shows equivalent `full cycles', defined as the charge throughput divided by twice the battery capacity.
		
		Figures \ref{degradation}a and \ref{degradation}b show the results for the simple linear degradation formula coupled with the bucket battery model (\ref{BM_degrad}). As only energy throughput and the maximum power level are taken into account in this model, no calendar ageing is predicted. Cycle ageing is deliberately underestimated because there is no temperature dependency in the model and experiments were done at 45$^{\circ}$C while the model was run at 25$^{\circ}$C.
		
		Figures \ref{degradation}c and \ref{degradation}d show the degradation predicted by the empirical degradation formula  from Schmalstieg \cite{Schmalstieg2014} coupled with the equivalent circuit battery model (\ref{ECM_degrad}). This formula was designed for the same battery chemistry (NMC), explaining the good correspondence for calendar ageing. The cycle ageing is less well predicted, since it was designed for different testing conditions. For example, the formula predicts very similar degradation for all cycles because it uses the mean voltage of the cycle. Since all cycles are centred around 50\% state of charge, their mean voltages are very similar. This illustrates that empirical correlations are not valid for operating conditions other than the ones tested. The initial degradation in particular is overestimated. This would result in the battery not being used at all, and therefore $\alpha$ and $\beta$ from (\ref{ECM_degrad}) were divided by 5 for the optimisation. By using this factor, the cycle degradation for the first 1000 cycles had the correct order of magnitude. 
		
		Figures \ref{degradation}e and \ref{degradation}f show the predictions according to the SEI-growth model coupled with the single particle model. Because the SPM models only one of the many physical processes responsible for li-ion battery degradation, the predictions are not fully accurate (e.g. degradation at low SoC is underestimated). However, the major trends are predicted better than the other two models.

		Table \ref{accuracy} summarises the accuracy of the three models. They were tested against data from the Mat4Bat project, as shown in Figure \ref{degradation}, and the data reported by Schmalstieg et al.\ \cite{Schmalstieg2014}. This allows a fair comparison between the single particle model, which is calibrated for the Mat4Bat data, and the equivalent circuit model, which is calibrated for Schmalstieg’s data. The equivalent circuit model is quite accurate except compared with the Mat4Bat cycle data. As noted above, this is due to extrapolation of the experimental outcomes without theoretical basis. The single particle model has the lowest average root mean square error in battery capacity prediction, which decreased from 11\% (of initial capacity) for the bucket model to 5\% (of initial capacity) for the single particle model.
		
		\begin{table}
			\begin{center}
				\caption{Root mean square error [\%] for the simulations compared with two degradation data sets for various battery models and the average error}
				\label{accuracy}
				\begin{tabular}{p{3cm} | c  c | c  c | c }
					& \multicolumn{2}{c|}{Calendar ageing} & \multicolumn{2}{c|}{Cycle ageing} & \\
					& Mat4Bat & Schmalstieg & Mat4Bat & Schmalstieg &  Average\\
					\hline
					Bucket				& 10.16 & n/a & 12.24 & n/a & 11.2 \\
					Equivalent circuit	& 3.24 & 3.77 & 12.23 & 3.14 & 5.60 \\
					Single particle		& 3.28 & 4.18 & 6.88  & 5.34 & 4.92 \\
				\end{tabular}
			\end{center}
		\end{table}
		
	\subsection{Optimal battery utilisation for arbitrage}
\begin{figure}
			\centering
			\includegraphics[width=14cm]{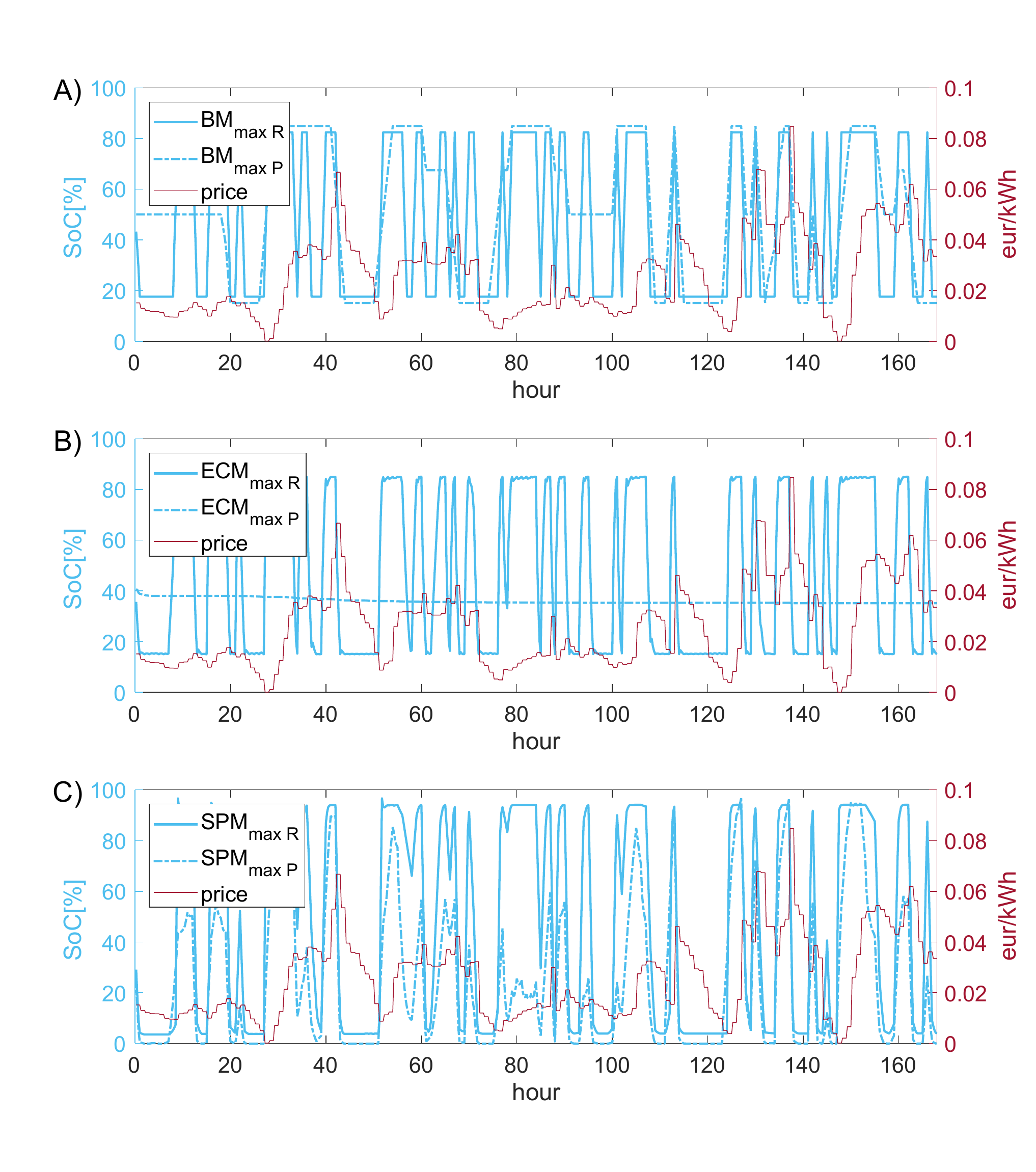}
			\caption{Wholesale price of electricity (right axis) and battery state of charge for the optimal battery utilisation (left axis) for price arbitrage in the first week of the year. Solid lines: maximising revenue (R); dashed lines: maximising profit (P). A) bucket model; B) equivalent circuit model; C) single particle model }
			\label{SoC_price_week}
		\end{figure}

		Having established a comparison between the degradation modelling approaches and experimental data, we now turn to optimal control of grid-connected batteries. The optimisation was carried out twice for each battery model, once to maximise the revenue, and in the second case to maximise total profit (revenue minus degradation cost), in order to assess the impact of co-optimising for degradation. These six optimal control profiles were validated with the single particle model as described in section \ref{postProcess}. A close-up of the first week of each validated optimal profile is shown in Figure \ref{SoC_price_week}. The profiles of the three battery models are given in the three subplots. The solid lines show the revenue-maximising outcomes while the dashed lines show the profit-maximising outcomes. In Figure \ref{SoC_price_week}, two main trends can be identified: the accessible capacity increases with increasing accuracy of the battery model, and battery utilisation decreases by co-optimising for degradation.
		
		The bucket battery model can access about 65\% of the battery capacity. This is because the optimal battery power has been rescaled by a factor of 0.65 to produce a safe current. The equivalent circuit model can access a larger proportion of the battery because it approximates diffusion and the (estimated) voltage is explicitly taken into account. The single particle model can access all the capacity because the same battery model is used for optimisation and validation. Still, the state of charge does not reach 100\% because the maximum capacity was rated at C/25 current (i.e.\ a discharge of 25 hours), and in practice not all this capacity can be accessed. 
		
		When degradation is taken into account (dashed lines), the battery is used less. Depending on the formula used to estimate the degradation, the optimal profiles will use the battery less in different ways. The bucket degradation model minimises the energy throughput and the maximal power. The equivalent circuit degradation model will push the SoC as close as possible to the SoC that minimises calendar ageing and minimise the charge throughput. The single particle degradation model will decrease the SoC and avoid high-power charging. As mentioned before, the empirical degradation formula identified by Schmalstieg \cite{Schmalstieg2014} and used with the equivalent circuit model, overestimates initial degradation because of the square root dependency on charge throughput. Even if it was decreased by a factor of 5, it still overestimated the degradation cost, reducing battery utilisation far more than the `real' optimum, to almost nothing.
		
\begin{figure}
			\centering
			\includegraphics[width=14cm]{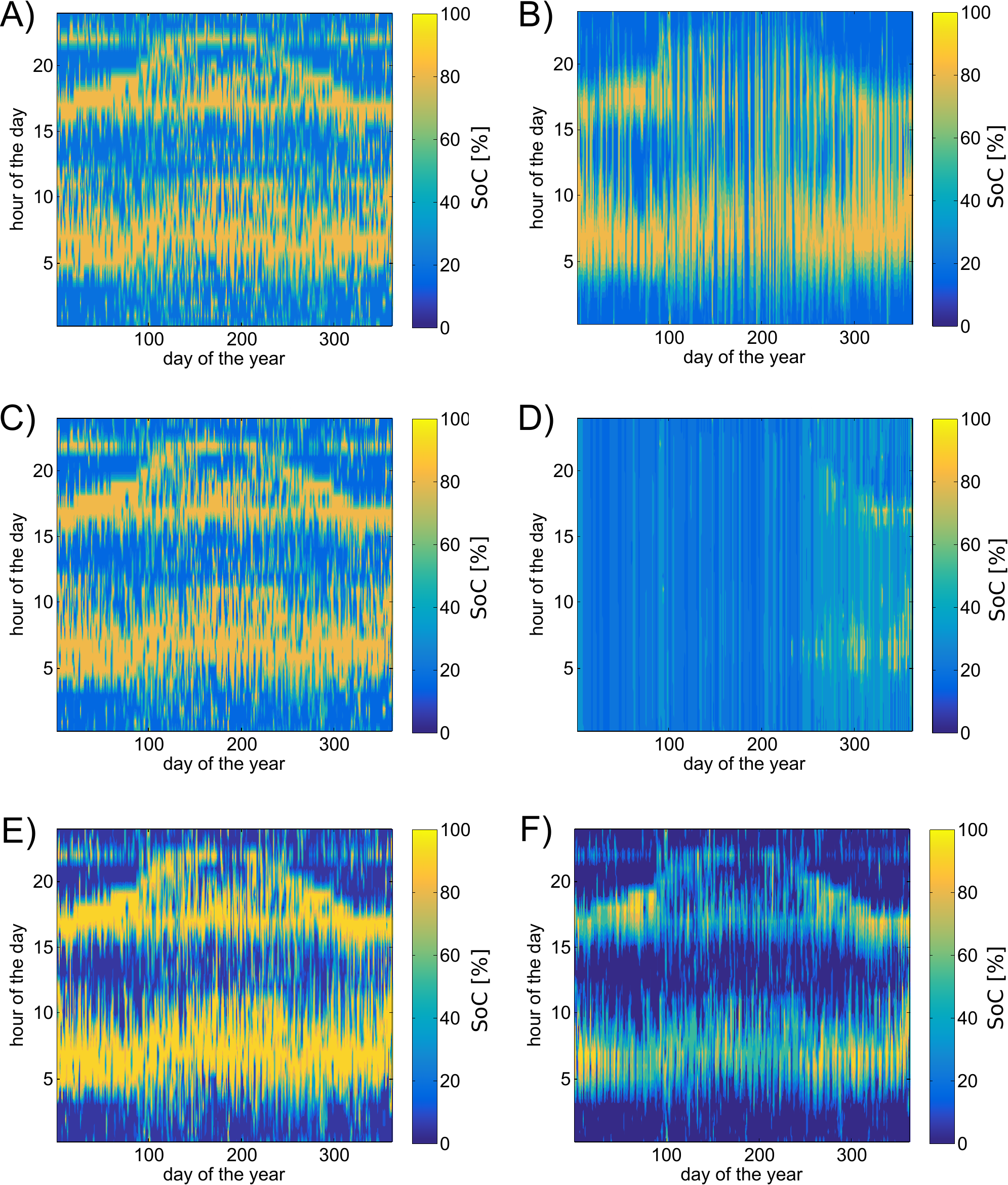}
			\caption{Battery SoC for every optimisation. Left: maximising revenue. Right: maximising profit. A) \& B) bucket model; C) \& D) equivalent circuit model; E) \& F) single particle model }
			\label{SoC_price_year}
		\end{figure}

		Figure \ref{SoC_price_year} shows the battery state of charge at every hour of every day for each optimisation. The same two trends can be identified as on Figure \ref{SoC_price_week}: more capacity can be accessed by more accurate models and optimising profit reduces utilisation. In Figure 6d, showing the result for the profit maximising with the equivalent circuit model, it can be seen that the battery utilisation increases later in the year, when the predicted degradation curve starts to flatten, decreasing the predicted degradation and hence increasing the battery utilisation. 
		
\begin{figure}
			\centering
			\includegraphics[width=9cm]{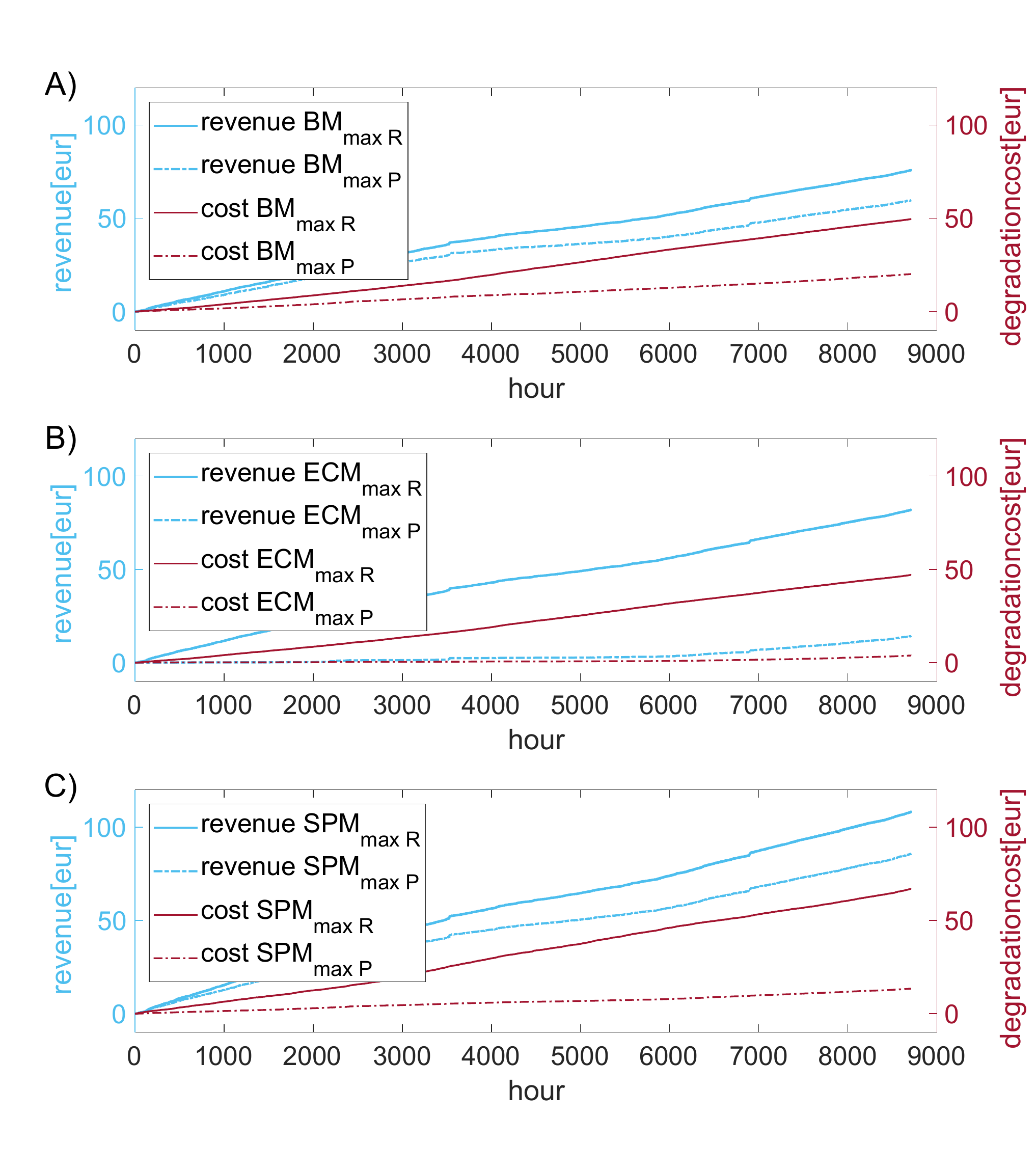}
			\caption{Optimal cumulative degradation cost (right axis) and cumulative revenue (left axis) for price arbitrage over the whole year. Solid lines: maximising revenue (R); dashed lines: maximising profit (P). A) bucket model; B) equivalent circuit model; C) single particle model}
			\label{Revenue_Degradation_year}
		\end{figure}

		Figure \ref{Revenue_Degradation_year} shows the cumulative revenue of each of the outcomes over the whole year in blue and the corresponding cumulative degradation cost (as predicted by the single particle model) in red. It indicates the impact both trends have on the revenue and degradation cost. The bucket model accounting for some degradation performs quite well. This is mainly a side-effect of the inability to access the full capacity because the estimated degradation (SEI growth) happens most at high SoC, and the model cannot access the high SoC region. This also decreases the revenue. 
		
		The equivalent circuit model can access more capacity, increasing the revenue. However, in the case of maximising profit, overestimated degradation (even if divided by a factor of 5) highly reduced the optimal utilisation, decreasing revenue (and degradation) to low levels. This is discussed in the next section. The single particle model can access the full battery, increasing revenue. Even when degradation is taken into account, and battery utilisation is reduced, the revenue stays high.
		
		Table \ref{overview} summarises the performance of the 6 optimisation scenarios. It shows the revenue, degradation cost and profit at the end of the year, as well as the relative capacity the battery has lost during the year as predicted by the single particle model. As expected, the lost capacity is lower if degradation is included in the optimisation. For the bucket and single particle battery models, this also increases profit. The equivalent circuit battery model with the unrealistic empirical degradation formula reduced battery utilisation too much, reducing the total profit. The profit of the single particle model accounting for degradation, 72.41 \euro{}, is a 175\% improvement over the profit of the simplest model (bucket model not accounting for degradation). Although the simulated degradation might not exactly correspond to actual degradation (see Figure \ref{degradation}), major improvements can still be expected by accounting for degradation.
		
		\begin{table}
			\begin{center}
				\caption{Revenue, degradation cost, profit and lost capacity at the end of the year for each optimisation and each battery model}
				\label{overview}
				\begin{tabular}{p{2.8cm} | Z{1cm}  Z{1cm}  Z{1cm}  Z{2cm} |  Z{1cm}  Z{1cm}  Z{1cm}  Z{2cm} |}
					& \multicolumn{4}{c|}{Max Revenue} & \multicolumn{4}{c|}{Max Profit} \\
					& Revenue [\euro{}] & Cost [\euro{}] & Profit [\euro{}] & Lost capacity [\%] & Revenue [\euro{}] & Cost [\euro{}] & Profit [\euro{}] & Lost capacity [\%] \\
					\hline
					Bucket				& 75.93 & 49.60 & 26.32 & 2.04 & 59.77 & 20.14 & 39.63 & 0.83\\
					Equivalent circuit	& 82.04 & 47.02 & 35.02 & 1.93 & 14.27 & 3.81  & 10.47 & 0.16\\
					Single particle		& 108.35& 66.95 & 41.40 & 2.75 & 85.82 & 13.42 & 72.41 & 0.55\\
				\end{tabular}
			\end{center}
		\end{table}
		
\section{Discussion}
Two important limitations arise from using the single particle model for the validation. The estimated lost capacity from Figure \ref{Revenue_Degradation_year} and Table \ref{overview} might be slightly wrong because only one degradation mode is included in the single particle model while there are many others (so degradation at low SoC, high power levels, low temperatures, etc.\ will be underestimated). Secondly, the single particle model gets an artificial advantage over the other models because the same model is used for optimisation and validation, while the other optimisations have a different model for optimisation versus validation. But given the relatively good accuracy of the model (Figure \ref{degradation}) against a large real experimental degradation data set, and the large performance difference, the general trends will most probably be valid in reality.

Because our optimisation does not consider the entire time horizon (see section \ref{slidingWindowParagraph}), the two-day optimisation time horizon limits long-term effects e.g.\ decreasing future revenue as the capacity available in the future decreases due to the current battery capacity fade. Especially for the equivalent circuit model, this is problematic. The empirical degradation formula predicts a high initial degradation, which flattens out later in the battery life. Counter-intuitively, this means that using the battery early in its life has a positive impact by reducing the degradation rate later in its life. But the 2-day window is too short to capture this effect, leading to suboptimal outcomes. Therefore, the equivalent circuit model probably performs better if a whole year is optimised at once because then this positive effect can be captured.

The decrease in battery degradation by maximising profit instead of revenue is about 80\% for the single particle optimisation, which is in line with other reported results. Pathak et al.\ \cite{Pathak2017} optimised an electrochemical battery model for a short period of time and for another application (fast charging). They repeated the same pattern hundreds of times and found that degradation could be decreased by well over 50\%. Patsios et al.\ \cite{Patsios2016} used a single particle model in a peak-shaving application. By running multiple (manually set) battery controls, they concluded battery degradation can be reduced by over 40\%. A full-scale optimisation should outperform this heuristic approach, so a reduction of 80\% seems realistic. 

\section{Conclusions}
In this paper, three dynamic battery and degradation models with increasing complexity were described, along with three different ways to model battery degradation. The accuracy of the models was tested by comparing them with a large degradation data set. The increase in model complexity improved overall accuracy, reducing the root mean square error in predicted capacity fade from 11\% to 5\%.

An optimal control problem was designed for a realistic scenario where a storage operator uses a lithium-ion battery for price arbitrage on the wholesale electricity market. It was assumed all prices were known a priori. The optimal battery utilisation profiles were validated using the single particle model and safety margins were introduced to avoid over- or under-charge.

It was found that, the more accurate the battery model, the more battery capacity could be used due to decreasing safety margins, increasing revenue by up to 42\%. More complex degradation models decrease degradation by up to 80\%, but this might decrease the overall profit by reducing the battery utilisation too much because degradation is overestimated in some scenarios. It is crucial to include all the necessary terms in the degradation model because the optimisation only minimises the terms explicitly present. 

The more complex and accurate models are nonlinear and require small time steps for stability, posing significant computational complexity challenges for the optimal control problem. The models have to be designed with optimisation in mind. In the case of the single particle model, Chebyshev discretisation was used instead of finite differences to reduce the number of state variables. If simultaneous discretisation in space and time is used, multiple shooting has to be used to reduce the overall number of optimisation variables. Even then, the optimisation still involves several thousands of variables. 

The total simulated profit increased by 175\% from the simplest to the most complex model. Real battery degradation might differ from the simulated degradation here, but still major improvements can be expected by increasing model accuracy. This clearly illustrates that techno-economic assessments for grid-connected storage could under- or over-estimate the batteries' potential, depending on how they account for safety margins in their sizing approach, and miss important issues such as degradation, because they almost exclusively use the simplest and least accurate battery models.

♥\section*{Acknowledgements}
This work was supported by VITO and EnergyVille, Belgium. The research leading to these results has been performed within the MAT4BAT project (http://mat4bat.eu/) and received funding from the European Union Seventh Framework Programme (FP7/2007-2013) under grant agreement nr 608931 and  under the SolSThore project which receives the support of the European Union, the European Regional Development Fund ERDF, Flanders Innovation \& Entrepreneurship and the Province of Limburg. Sina Ober-Blöbaum was supported by the EPSRC project: "Fractional Variational Integration and Optimal Control"; ref: EP/P020402/1.

\section*{References}

\bibliographystyle{elsarticle-num}
\bibliography{References}

\appendix
\section{Model equations of the single particle model} \label{appendix_SPM}

\begin{tabular}{p{4cm} p{13cm}}
	Arrhenius relation for diffusion constants \cite{Guo2011} & \begin{equation} D_i=D_i^{\text{ref}} \text{exp}\left[ \frac{E_{D,i}}{R}  \left( \frac{1}{T(t)} - \frac{1}{T^{\text{ref}}} \right) \right] \end{equation} \\
	
	Arrhenius relation for rate constants \cite{Guo2011} & \begin{equation} k_i=k_i^{\text{ref}} \text{exp}\left[ \frac{E_{k,i}}{R}  \left( \frac{1}{T(t)} - \frac{1}{T^{\text{ref}}} \right) \right] \end{equation} \\
	
	Exchange current density at electrode i \cite{Ning2004} & \begin{equation} j_{i,0} = nFk_ic_i(R_i,t)^\alpha c_{ \text{el}}^{1-\alpha} \left( c_i^{ \text{max}}-c_i(R_i,t) \right)^{1-\alpha} \end{equation} \\
	
	Bulter-Volmer reaction kinetics at electrode i \cite{Ning2004} & \begin{equation}  J_i = j_{i,0} \left( \text{exp} \left( -\frac{\alpha nF}{RT(t)} \eta_i \right) -  \text{exp} \left( \frac{(1-\alpha) nF}{RT(t)} \eta_i \right) \right)  \end{equation}\\
	
	SEI side reaction current density \cite{Christensen2005} & \begin{equation}  i_{ \text{sei}} = \frac{\text{exp} \left( -\frac{nF}{RT(t)} \eta_{ \text{neg}} \right)}{ \frac{1}{n_{ \text{sei}}Fk_{ \text{sei}}\text{exp} \left( -\frac{n_{ \text{sei}}F}{RT(t)}(OCV_{ \text{neg}}-0.4) \right)} + \frac{\delta (t)}{n_{ \text{sei}}FD_{ \text{sei}}}}   \end{equation}\\
	
	Open circuit voltage & \begin{equation}  OCV_i^{\text{ref}} = f \left( c_i(R_i,t) \right)  \end{equation} \\
	
	Battery voltage & \begin{equation}\begin{aligned}  V(t) =\: &OCV_{ \text{pos}}^{\text{ref}} -  OCV_{ \text{neg}}^{\text{ref}} + \left( T(t)-T^{\text{ref}} \right) \frac{\partial OCV}{\partial T} \\  &- \left( \eta_{ \text{neg}}-\eta_{ \text{pos}} \right) - \left( R_{ \text{batt}} + r_{ \text{sei}}\delta (t) \right)I(t)  \end{aligned} \end{equation} \\
	
	Thermal model \cite{Guo2011} & \begin{equation} \begin{aligned} \rho v c_p \frac{\partial T(t)}{\partial t} = \: &I(t)^2 R_{ \text{batt}} + I(t) \left( \eta_{ \text{neg}} - \eta_{ \text{pos}} \right) \\ & + I(t)T(t)\frac{\partial OCV}{\partial T} - hA_{ \text{batt}} \left( T(t)-T_{ \text{env}} \right) \end{aligned} \end{equation} \\
	
\end{tabular}
\begin{tabular}{p{4cm} p{13cm}}
	
	SEI layer growth \cite{Ning2004} & \begin{equation}  \frac{\partial \delta (t)}{\partial t} = \frac{i_{ \text{sei}}M}{n_{ \text{sei}}F\rho_{ \text{sei}}}   \end{equation}\\
	
	Loss of lithium & \begin{equation}  \frac{\partial L(t)}{\partial t} = i_{ \text{sei}}A_n  \end{equation}\\

	Diffusion in positive electrode \cite{Ning2004} & \begin{equation}  \frac{\partial c_p(r,t)}{\partial t} = \frac{D_p}{r^2}\frac{\partial}{\partial r} \left( r^2\frac{\partial c_p(r,t)}{\partial r} \right)  \end{equation}\\
	
	Diffusion in negative electrode \cite{Ning2004} & \begin{equation}  \frac{\partial c_n(r,t)}{\partial t} = \frac{D_n}{r^2}\frac{\partial}{\partial r} \left( r^2\frac{\partial c_n(r,t)}{\partial r} \right) - \frac{i_{ \text{sei}}a_n}{nF}  \end{equation}\\
\end{tabular}

\begin{tabular}{p{1.5cm} p{12.5cm} p{2cm}}
	$\alpha$ 	& Charge transfer coefficient 								& 			\\
	$\delta(t)$ & Thickness of the SEI layer at time t 						& $m$		\\
	$ \frac{\partial OCV}{\partial T} $  & Entropic coefficient of the open circuit voltage	& $V\ K^{-1}$ \\
	$ \eta_i$ 	& Chemical reaction overpotential at electrode i or SEI side reaction& $V$	\\
	$ \rho$ 	& Density of the battery									& $kg\ m^{-3}$ \\
	$ \rho_{ \text{sei}}$ & Density of the SEI layer								& $kg\ m^{-3}$ \\
	$ A_i$ 		& Electrode surface of electrode i 							& $m^2$ 	\\
	$ A_{ \text{batt}}$ & Total battery surface										& $m^2$ 	\\
	$ a_i$ 		& Specific electrode surface of electrode i 				& $m^2\ m^{-3}$	\\
	$ c_{ \text{el}}$ 	& Lithium concentration in the electrolyte					& $mol\ m^{-3}$ \\
	$ c_i(r,t)$ & Lithium concentration in electrode i at radius r and time t & $mol\ m^{-3}$ \\
	$ c_{i}^{ \text{max}}$ & Maximum lithium concentration in electrode i			& $mol\ m^{-3}$ \\
	$ c_p$ 		& Heat capacity of the battery								& $J\ kg^{-1}K^{-1}$ 		\\
	$ D_i$ 		& Diffusion coefficient of Li in electrode i or SEI layer, temperature dependent																& $m^2\ s^{-1}$\\
	$ D_i^{\text{ref}}$ & Diffusion coefficient of Li in electrode i or SEI layer at reference temperature																& $m^2\ s^{-1}$ \\
	$ E_{D,i}$ & Activation energy of the Arrhenius relationship for the diffusion coefficient at electrode i																& $J\ mol^{-1}$ \\
	$ E_{k,i}$ & Activation energy of the Arrhenius relationship for the rate coefficient at electrode i																		& $J\ mol^{-1}$ \\
	$ F $ 		& Faraday constant											& $C\ mol^{-1}$	\\
	$ h $ 		& Convective heat transfer coefficient						& $W\ m^{-2}K^{-1}$\\
	$I(t)$ 	& Total battery current at time t 							& $A$ 		\\
	$i_{ \text{sei}}(t)$ & SEI side reaction current density						& $A\ m^{-2}$\\
	$J_i$		& Reaction current density in electrode i					& $A\ m^{-2}$\\
	$J_{i,0}$	& Exchange current density in electrode i					& $A\ m^{-2}$\\
	$k_i$ 		& Chemical rate constant of the main reaction in electrode i or SEI side reaction, temperature dependent						& $m^{4-3\alpha}\ s^{-1}\ mol^{\alpha-1}$ \\
	$k_i^{\text{ref}}$	& Chemical rate constant in electrode i or SEI  side reaction at reference temperature						& $m^{4-3\alpha}\ s^{-1}\ mol^{\alpha-1}$ 			\\
	$L(t)$	& Lost lithium content at time t							& $Ah$ 		\\
	$M$			& Molar volume of SEI reaction products						& $kg\ mol^{-1}$\\
	$n$			& Number of electrons in the main reaction					& $-$		\\
	$n_{ \text{sei}}$	& Number of electrons in the SEI side reaction				& $-$		\\
	$OCV_i^{\text{ref}}$ & Open circuit voltage of electrode i at reference temperature& $V$ 	\\
	
\end{tabular}
\begin{tabular}{p{1.5cm} p{12.5cm} p{2cm}}
	
	$R$ 		& Ideal gas constant										& $J\ mol^{-1}K^{-1}$ \\ 
	$R_{ \text{batt}}$	& Total battery resistance									& $\Omega$ 	\\
	$R_i$ 		& Radius of the particle of electrode i						& $m$ 		\\
	$r_{ \text{sei}}$ 	& Resistance of the SEI layer per unit of SEI thickness		& $\Omega\ m^{-1}$\\
	$T(t)$ 		& Uniform battery temperature at time t						& $K$ 		\\
	$T_{ \text{env}}$ 	& Temperature of the environment							& $K$ 		\\
	$T^{\text{ref}}$ 	& Reference temperature										& $K$ 		\\
	$V(t)$ 		& Battery voltage at time t									& $V$ 		\\
	$v$ 		& Volume of the battery										& $m^3$		
\end{tabular}

\section{Post processing without safety margins} \label{appendix_postProcess}
In section \ref{postProcess}, it was explained how safety margins were introduced in the post processing to guarantee a current which wouldn't violate the voltage constraints. This decreased the usable capacities for the batteries, especially for the ones controlled by the bucket and equivalent circuit models. This appendix introduces a second way of avoiding the over- and under-voltages without decreasing the usable capacity.

In the validation phase, a battery of the same size (without safety margins) is simulated using the single particle model. It tries to follow the optimal profile but when it reaches one of the voltage limits, it deviates from the profile and instead keeps the voltage fixed in order to avoid over- or under-voltages. In this way, the full battery capacity can be used.

Table \ref{overview2} gives the numerical outcomes using this approach. Compared with Table \ref{overview}, the revenue increases, but because the battery operates at high state of charge more often, this also leads to significantly more battery degradation. For the cases where degradation is ignored, this dramatically reduces the profit, even resulting in an overall loss when the batteries are controlled by the bucket or equivalent circuit models. When the optimisation accounts for degradation and maximises the net profit, the overall profit remains almost the same. However for the bucket-model controlled battery this hides a 40\% increase in the revenue and a doubling of the degradation.

\begin{table}
	\begin{center}
		\caption{Revenue, degradation cost, profit and lost capacity at the end of the year for each optimisation and each battery model with a different post processing approach}
		\label{overview2}
		\begin{tabular}{p{2.8cm} | Z{1cm}  Z{1cm}  Z{1cm}  Z{2cm} |  Z{1cm}  Z{1cm}  Z{1cm}  Z{2cm} |}
			& \multicolumn{4}{c|}{Max Revenue} & \multicolumn{4}{c|}{Max Profit} \\
			& Revenue [\euro{}] & Cost [\euro{}] & Profit [\euro{}] & Lost capacity [\%] & Revenue [\euro{}] & Cost [\euro{}] & Profit [\euro{}] & Lost capacity [\%] \\
			\hline
			Bucket				& 97.71 & 123.38& -25.67& 5.07 & 82.41 & 43.30 & 39.11 & 1.78\\
			Equivalent circuit	& 104.81& 135.28& -30.47& 5.56 & 18.95 & 6.12  & 12.83 & 0.25\\
			Single particle		& 116.56& 95.41 & 21.15 & 3.92 & 85.84 & 13.53 & 72.31 & 0.56\\
		\end{tabular}
	\end{center}
\end{table}

\end{document}